\documentclass[12pt]{iopart}

\usepackage{graphicx}
\begin{document}

\title{The background from single electromagnetic subcascades for a stereo system of air Cherenkov telescopes}

\author{Dorota Sobczy\'{n}ska}

\address{University of {\L }\'{o}d\'{z}, 
Department of Astrophysics, Pomorska 149/153, 90-236 {\L }\'{o}d\'{z},Poland}
\ead{ds@kfd2.phys.uni.lodz.pl}

\begin{abstract}

The MAGIC experiment, a very large Imaging Air Cherenkov Telescope (IACT) with sensitivity to low energy (E $<$ 100 GeV) VHE gamma rays, has been operated since 2004. It has been found that the $\gamma$/hadron separation in IACTs becomes much more difficult below 100 GeV \cite{crab2008}.\\ 
A system of two large telescopes may eventually be triggered by hadronic events containing Cherenkov light from only one electromagnetic subcascade or two $\gamma$ subcascades, which are products of the single $\pi^0$ decay. This is a possible reason for the deterioration of the experiment's sensitivity below 100 GeV. In this paper a system of two MAGIC telescopes working in stereoscopic mode is studied using Monte Carlo simulations.\\ 
The detected images have similar shapes to that of primary $\gamma$-rays and they have small sizes (mainly below 400 photoelectrons (p.e.)) which correspond to an energy of primary $\gamma$-rays below 100 GeV.
The background from single or two electromagnetic subcascdes is concentrated at energies below 200 GeV. 
Finally the number of background events is compared to the number of VHE $\gamma$-ray excess events from the Crab Nebula.
The investigated background survives simple cuts for sizes below 250 p.e. and thus the experiment's sensitivity deteriorates at lower energies.

\end{abstract}

\submitto{\JPG}
\maketitle

\section{Introduction}

The Imaging Air Cherenkov technique \cite{hinton} is successfully used since the detection of the first TeV $\gamma$-ray source (the Crab Nebula) by the Whipple collaboration \cite{whipple}. The telescopes measure Cherenkov light of Extensive Air Showers (EAS), which is recorded as an image in the camera of the telescope together with the light from the Night Sky Background (NSB). In order to analyze the image a so-called cleaning procedure is used to remove NSB-dominated pixels below a certain threshold of the signal-dominated pixels.

The number of events induced by the hadronic background is several orders of magnitude larger than the number of the registered $\gamma$-rays from the source. A method to select $\gamma$-rays out of a hadron dominated event sample was first proposed by Hillas in 1985 \cite{hillas}. This method is based on a parameterization of the shape and the orientation of the image. Parameters relevant for this study are: WIDTH (RMS of the signal along minor axis of the image), LENGTH (RMS of the signal along major axis of the image), ALPHA (angle between the major axis of the image and the line between the center of the image and the source position in the camera), $\theta^2$ (the square of the distance between the source position and the intersection of the major axes - for a stereo observation) and SIZE (sum of all signals from pixels contained in the image). A narrow shape indicates a $\gamma$-ray as primary particle; the direction of the image main axis determines the direction of the primary particle and thus points towards the center of the camera (if the telescope is directed towards a source of $\gamma$- ray emission), while the hadronic background is isotropically distributed.\\

In order to lower the energy threshold and improve the sensitivity of IACTs, stereoscopy is used and telescopes with larger mirror area have been built. The $\gamma$/hadron separation methods which are used now are more sophisticated (such as \cite{random2008,kraw2006}), but are still based upon the original Hillas parameters. The time structure of the image can be used in the data analysis as an additional parameter as it is shown in \cite{time2009}.
Recent experiments with large reflectors like CANGAROO \cite{enom2002,mori}, HESS \cite{wasil2005,hofmann}, MAGIC \cite{baix2004,mariotti}, VERITAS \cite{weekes2002,weekes} made a major contribution to the very fast development of ground-based $\gamma$-ray astronomy (see figure 1 in \cite{hinton2007}).\\
The detection of low energy events is possible by using a very large telescope, which requires a parabolic shape of the main reflector dish (like that in MAGIC) to avoid broadening of the time profile of the Cherenkov signal. This causes non-negligible off-axis aberrations which is one of the reasons for the worsening of the $\gamma$-ray selection efficiency. In the low energy region the $\gamma$/hadron separation becomes more difficult due to higher relative fluctuations in the shower development, which results in larger fluctuations of the Cherenkov light density \cite{bhat,sob2009} and image parameters. The image parameters also depend upon the  local azimuth and the elevation of the $\gamma$-ray  trajectory because of  the influence of the Earth's Magnetic Field (EMF) on the images \cite{comm08}. The $\gamma$-ray images become wider and their major axis  changes the direction a bit  because of the EMF influence on the directions of charged particles in the shower. Both effects make  $\gamma$-ray images more similar to the background events and finally the efficiency of the $\gamma$/hadron separation become worse.\\ 
When the image contains photons from only a single charged particle (for example muons), it may have a very similar shape as the $\gamma$ cascade image and a very narrow arrival time distribution \cite{mirzoyan}. It has been shown in \cite{mirzoyan} that this background can be significantly rejected by using a Fast Analog Digital Converter (FADC) in the readout system but at the moment selecting single distant muons only from their time spread is not possible \cite{time2009} in the MAGIC experiment.\\
There is another kind of background: images of one electromagnetic subcascade from hadron-initiated showers \cite{sob2007}. Telescopes may be triggered by light produced by $e^+$ and $e^-$ from this electromagnetic subcascade. Those events are called {\it false $\gamma$ images} in this paper. Electromagnetic subcascades appear in the hadronic shower mostly as products of the $\pi^0$ or $\eta$ decay process. The image may contain photons from only two $\gamma$ subcascades, which originated from the same particle. I shall call them {\it one $\pi^0$ images}.
There is no physical reason for differences in shapes of the false and the true $\gamma$-ray images except for the different height of the first $e^+$, $e^-$ pair production. The false $\gamma$ events start deeper in the atmosphere, thus a narrower angular distribution of the charged particles is expected. False $\gamma$-ray images are therefore narrower than real $\gamma$-ray events.  
In the case of one $\pi^0$ events one may expect slightly wider images due to the existence of the separation angle between the decay products.
The orientation of the major image axis of the hadronic background is randomly distributed and false $\gamma$ and one  $\pi^0$ images should have the same features. One can expect that the parameter which is describing the orientation of the image is still a good quantity for the true $\gamma$-ray selection. 
In \cite{sob2007} the fraction of false $\gamma$ and one $\pi^0$ events  in the proton showers has been estimated and discussed for a single telescope like MAGIC. 
As it is shown in \cite{maier} the background surviving a $\gamma$-ray selection for four telescopes (working in stereo mode) of the VERITAS experiment is explained as showers which transfer a large fraction of the primary energy to electromagnetic subshowers during the first few interactions. This result confirms the fact that single electromagnetic subcascade images are a hardly reducible background in the stereo observations.\\
The results presented in this paper are based on a Monte Carlo simulation for two Cherenkov telescopes similar to the MAGIC II experiment \cite{carmo2007}. 
In the following I present the MC study and estimate the fraction of false $\gamma$ and one $\pi^0$ events  in the proton showers for four different trigger thresholds. The fraction of this background in the total sample of triggered proton events is also calculated for different SIZE bins. The primary energy of true $\gamma$-rays which corresponds to a chosen SIZE range was found. The distribution of the parameters describing the image shape are presented to show the similarity of true and false $\gamma$ events. The ratio of the expected number of both false $\gamma$ and one $\pi^0$ images to that of primary $\gamma$-rays from the Crab Nebula direction is calculated in different SIZE bins and for different trigger thresholds.  This ratio is shown in two cases. In the first case no $\gamma$/hadron separation method has been applied to simulations, in the second case simple width and length cuts were taken into account. The occurrence of false $\gamma$ and one $\pi^0$ events is an important reason for the suppressed $\gamma$/hadron separation efficiency in the low SIZE range of the stereo IACTs.   
\section{Monte Carlo simulations}

MAGIC II \cite{carmo2007} was chosen as an example of a stereo system of IACTs. The system contains two 17m diameter telescopes. The distance between them is 85 m. The first  MAGIC telescope \cite{baix2004,albert} started operation at the end of 2003, while the second one is in its commissioning phase (2009). The experiment is located on the Canary Island of La Palma (28.8N, 17.9W) at the Roque de los Muchachos Observatory, 2200 m above sea level. 
The telescopes have a parabolical shape with a focal length of 17 m. 
The diameter of each telescope's mirror dish is 17 m. The total reflector areas cover more than  230 $m^2$ each. The inner part of the MAGIC I camera consists of 397 photomultipliers (PMT) with 0.1$^o$ diameter.
An additional 180 PMTs with 0.2$^o$ diameter form the outer part of the camera. The hexagonal shaped camera covers in total almost 4$^o$. 325 pixels (in the inner part of the camera) are used for the trigger. The trigger requires a time coincidence and nearest neighbour logic. In the simulations presented in this paper the second telescope is exactly the same as the first one.

The shower development in the atmosphere was simulated using the CORSIKA code (version 6.023 \cite{heck,knapp}) with GHEISHA and VENUS as low (primary momentum below 80 GeV/c) and high energy interaction models, respectively. Showers initiated by primary protons with energies between 30 GeV and 1 TeV following a power law with a differential spectral index of -2.75 were simulated. The impact parameter was distributed randomly within a circle (radius of 1.2 km) around the center of the telescope system.
The showers were simulated within a cone with an opening angle of 5.5$^o$ at a zenith angle of 20$^o$ and an azimuth of 0$^o$ (showers directed to the north).
$25 \times 10^6$ primary proton events were simulated. The coordinates of the Earth's Magnetic Field have been chosen as 29.5 $\mu T$ in the north direction and 23.0 $\mu T$ in vertical direction (down).\\    
For the $\gamma$ cascade simulations the impact parameter was randomly distributed within a circle of 350 m radius with an energy range of 10 GeV to 1 TeV. The density of the Cherenkov light (for primary $\gamma$-rays) decreases very fast for impact parameters larger than approximately 120 m. Beyond 300 m  the probability to trigger both telescopes is negligible. The differential spectral index was chosen to be -2.6 (which is the index of the Crab spectrum for energies above 300 GeV \cite{wagner} or 500 GeV \cite{aha2004}). The direction of $0.5 \times 10^6$ simulated $\gamma$ cascades was fixed to a zenith angle of 20$^o$ and an azimuth angle of 0$^o$ (parallel to the optical axes of both telescopes).\\
Primary hadrons that are heavier than protons have not been simulated because the Cosmic Ray (CR) background is dominated by protons. Accordingly only proton induced showers are commonly used in IACT simulations. The simulation of a significant number of showers initiated by heavier nuclei requires very long computation time, which goes beyond the scope of this paper.

The CORSIKA code was incorporated to record additional information about each subcascade that has been produced in the EAS. The type of the charged particle, which was responsible for the creation of each Cherenkov photon is also saved in the output of the program.

The simulations were done for two telescopes working in a stereo system (coincidence is required for the trigger). The distance between the telescopes was chosen to be 85 m. Both IACTs had the same mirrors and cameras, which are similar to those of the MAGIC telescope \cite{albert,bario}. Rayleigh and Mie scattering of light in the atmosphere were taken into account according to the Sokolsky formula \cite{sokol}. The detector simulation consists of two parts: The first part (reflector) includes the full geometry of the mirrors and their imperfections. The camera properties, such as additional reflections in Winston cones, photocathode quantum efficiency \cite{bario} and its fluctuation, and the photoelectron collection efficiency were considered in the second program.\\
The night sky background (NSB) of $1.75  \times 10^{12} ph/(m^2~sr~s)$, which was measured on La Palma \cite{nsb}, was included in the simulation before checking the trigger conditions. Photoelectrons made by the NSB were not added to the images to avoid the necessity of the so called cleaning procedure. One may expect that an image cleaning with too high cleaning levels may make images artificially narrower. 
A comparison of the image parameters after cleaning may thus be less reliable. 
A single telescope was triggered by a shower if the output signals in three next neighbouring pixels (3 NN) exceed a certain threshold within a time window of 3 ns. 
The trigger thresholds have been chosen to be: 2, 3, 4, 5 in photoelectrons (hereafter referred to as {\it p.e.}), which corresponds to the signal of 2, 3, 4 and 5 photoelectrons (p.e.) arriving exactly at the same time \cite{sob}, respectively.\\

\section{Results and discussion}
\subsection{Fraction of the false $\gamma$ and one $\pi^0$ events in the total protonic background}
 The OFF parameter is defined as the angular distance between the telescope axis and the shower axis. Both OFF and the impact parameter were simulated in relatively large ranges in order to cover the area and the directions of all possible events that can trigger the IACT system. 
As mentioned in the Monte Carlo section, the impact parameter is defined as the distance between the shower core axis and the center of the telescope system. Figure 1a shows the dependence between the OFF parameter and the simulated impact parameter of proton events for a chosen trigger threshold of 3 p.e..
False $\gamma$ and one $\pi^0$ candidates are images with a small light contribution (less than 10$\%$ of the total SIZE) from the hadronic and muonic part of the shower. This kind of images ({\it electromagnetic events}) are presented in figure 1b. One $\pi^0$ and false $\gamma$ events are shown in figures 1c and 1d, respectively. Also in those cases the images are allowed to contain less than 10$\%$ of photoelectrons produced by other particles from shower. 
One $\pi^0$ events have an impact parameter mostly below 800 m, while false $\gamma$ events may trigger also with larger impact parameters. Similar results have been shown in \cite{sob2007}. Around 20$\%$ of the one $\pi^0$ events have impact parameters larger than 800 m. The fraction of false $\gamma$ events with impact parameters larger than 800m is around 20$\%$ for a stereo system of IACTs.\\
\begin{figure}
\begin{center}
\includegraphics*[width=14cm]{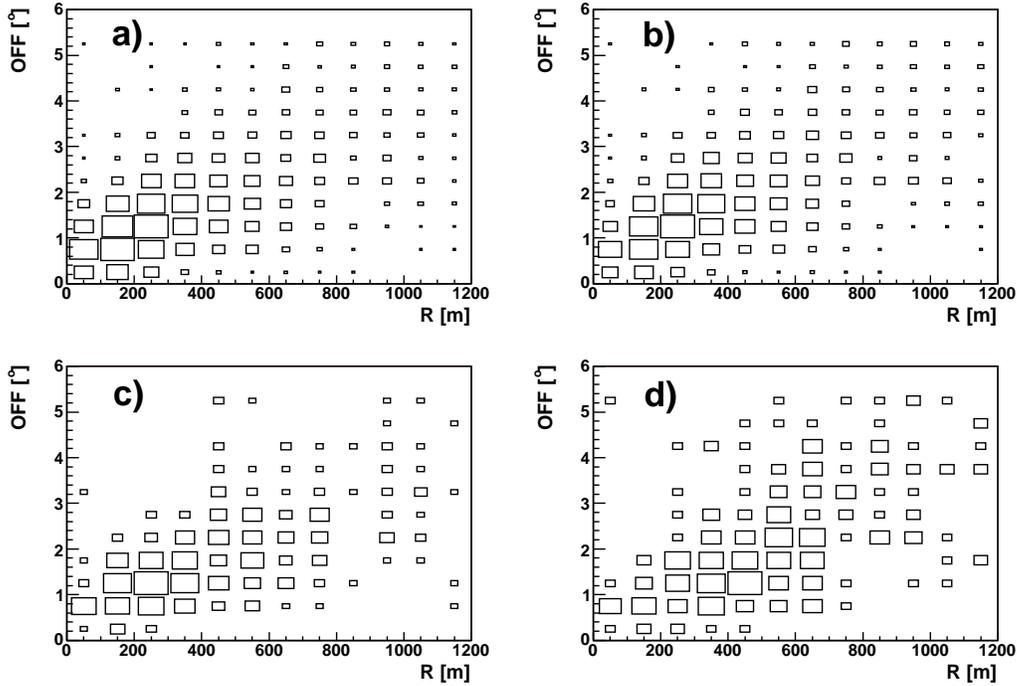}
\end{center}
\caption{Dependence of the OFF angle on the impact parameter for a trigger threshold of 3 p.e.: {\bf a)} all triggered proton showers (in total 2805 events; bin contents from 1 to 200 events); {\bf b)} electromagnetic events (in total 1913 events; bin contents from 1 to 142 events); {\bf c)} one $\pi^0$ events (in total 309 events; bin contents from 1 to 22 events); {\bf d)} false $\gamma$ events (in total 232 events; bin contents from 1 to 12 events). The area of a square is proportional to the number of events, but absolute numbers in each figure are different. Showers were simulated within a cone with an opening angle of 5.5 deg. The simulated energy spectrum is described in the section ``Monte Carlo Simulations''  }

\label{impact}
\end{figure} 
Figure 2 shows the primary energy distribution of the simulated and triggered proton events, for a trigger threshold of 3 p.e. The electromagnetic, one $\pi^0$ and false $\gamma$ events are also presented in this figure. The distributions for both one $\pi^0$ (dashed-dotted histogram) and false $\gamma$ (solid histogram) events are much steeper than that of all triggered events for primary energies above 100 GeV. Similar results have been shown in \cite{sob2007} for a single telescope and a different trigger condition. Approximately only 8$\%$ of all one $\pi^0$ events and only 5$\%$ of all false $\gamma$ events have an energy above 200 GeV, respectively.\\
\begin{figure}
\begin{center}
\includegraphics*[width=9cm]{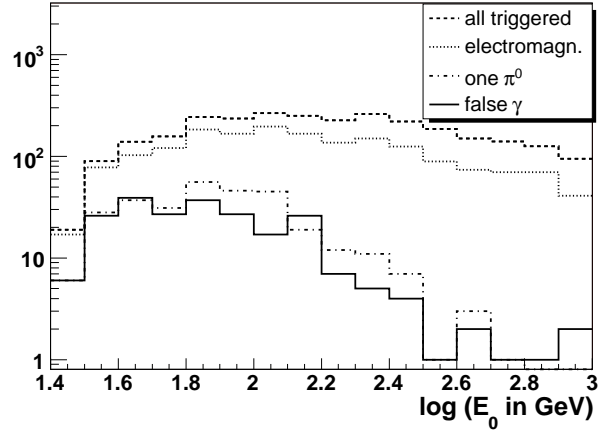}
\end{center}
\caption{Energy distribution for proton events (trigger threshold of 3 p.e.).}
\label{energy}
\end{figure} 
Assuming that the probability to trigger an event, that has an impact parameter larger than 1200 m or is inclined more than 5.5$^o$ with respect to the telescope axis, is negligible in comparison to all triggered showers, one may evaluate the expected number of triggered events for energies above 1 TeV. 
A simple power law fit of the energy distribution tail was used as an extrapolation function in order to estimate the number of showers above 1 TeV. 
For all the triggered events and the electromagnetic events, energies above 600 GeV were considered as the tail of the energy distribution. In case of false $\gamma$ and one $\pi^0$ events, the energy range between 100 GeV and 1 TeV was fitted.  The expected number of false $\gamma$ and one $\pi^0$ events has been estimated for images, which contain up to 10 $\%$ of light from other particles from the shower.
 The fraction of  electromagnetic, false $\gamma$ and one $\pi^0$ images in the expected protonic Cosmic Ray background was calculated for all simulated trigger thresholds. Figure 3 shows how these fractions are changing with the trigger threshold, respectively. 
The contribution of false $\gamma$ images in all triggered events decreases from 13$\%$ to 3$\%$ when the trigger threshold increases from 2 p.e. to 5 p.e. The fraction of one $\pi^0$ images changes from 12$\%$ to 6$\%$. 
The estimated fractions are lower (0.1 - 0.4 $\%$) for pure false $\gamma$ and one $\pi^0$ events. The pure false $\gamma$ or one $\pi^0$ events are defined as images, which  have no contribution from the rest of the shower.
 The fraction of pure electromagnetic events in all triggered events is reduced by 5$\%$ -  13$\%$ (compared to that of figure 3) for trigger thresholds from 2 p.e. to 5 p.e., respectively.\\
\begin{figure}
\begin{center}
\includegraphics*[width=9cm]{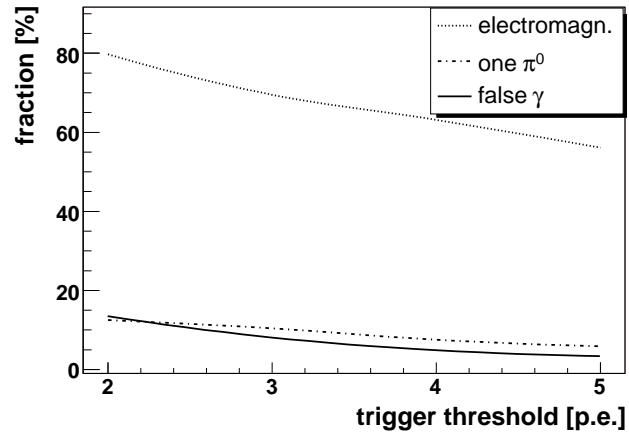}
\end{center}
\caption{Fraction of the interesting events in the total protonic background as a function of the trigger threshold.}
\label{fraction}
\end{figure} 
Hillas parameters have been calculated for images which have no contamination of the NSB. No cleaning procedure has been applied to these images. At first, the SIZE (the sum of all detected photoelectrons) was calculated. Figure 4 shows the SIZE distribution for a trigger threshold of 3 p.e., which was obtained for images registered by one of the telescopes. Both false $\gamma$ events (solid histogram in figure 4) and one $\pi^0$ images (dashed-dotted histogram) have small SIZES because their primary energy is low (see figure 2). Apart from that a relatively large fraction of the proton showers, which may imitate a real $\gamma$-ray, have impact parameters above 400 m, where the expected SIZE is low.\\ 
\begin{figure}
\begin{center}
\includegraphics*[width=9cm]{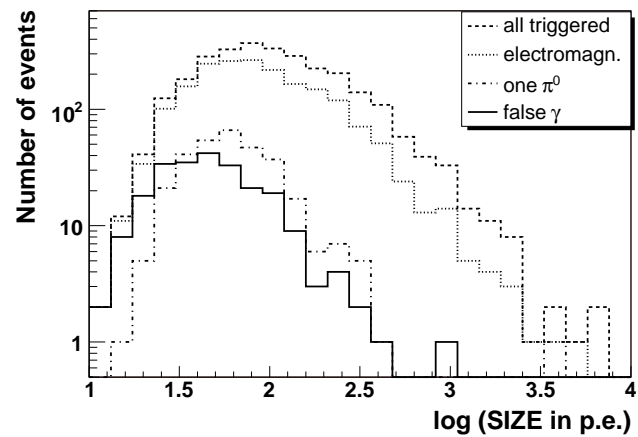}
\end{center}
\caption{SIZE distributions for proton showers; trigger threshold 3 p.e.}
\label{size}
\end{figure} 
The fraction of both false $\gamma$ and one $\pi^0$ events in the total number of triggered proton showers depends on the chosen SIZE interval. As the shower is measured simultaneously by two telescopes, one may require in the analysis one of the following four possible conditions: both SIZES (1), at least one SIZE (2), the SIZE measured by a fixed telescope (3) or average SIZE (4) are within the chosen interval. The ratio of the background from both false $\gamma$ and one $\pi^0$ event to the total sample of the triggered proton showers was calculated for five SIZE bins and all three conditions, respectively. The results are summarized in table 1 for a trigger threshold of 3 p.e. Additionally, the estimated energy of primary $\gamma$-rays for each SIZE interval is given in the last column. This energy was estimated as the peak position of the energy distribution of the primary $\gamma$-rays in the chosen SIZE interval (for condition (3) only).
While the numbers of triggered events that are fulfilling the conditions (1), (2) and (3) are different, the corresponding fractions (as presented in table 1) remain at a similar level.
 The expected background from both false $\gamma$ and one $\pi^0$ events is around 4$\%$ of the total background for an energy of true $\gamma$-rays around 100 GeV. It increases significantly at lower energies. All triggered proton showers (with SIZE below 100 p.e.) contain approximately 30$\%$ of events which are detected as false $\gamma$ or one $\pi^0$ events.\\
It is worth mentioning here that for a trigger threshold of 3 p.e., the estimated energy of primary $\gamma$-rays for each SIZE interval is the same as shown in \cite{sob2007} for a trigger threshold of 4 p.e. 
However the analysis in \cite{sob2007} has been done for a single telescope with a 4NN trigger logic condition. The trigger threshold of 3 p.e. for a system of IACTs has been chosen to show the expected fraction of events imitating real $\gamma$-rays with similar energies as it was discussed in \cite{sob2007}.

\begin{table}
\caption {\label{tab1}Fraction of events unrecognised by shape in the total sample of triggered proton showers for five SIZE bins (trigger threshold 3 p.e.). The last column corresponds to the position of the energy peak of $\gamma$-ray simulations.}

\begin{indented}
\item[]\begin{tabular}{@{}*{8}{lllllll}}
\br                              
SIZE &both telescopes&at least one&telescope 1& an average & energy peak\\
 & (1) & (2) & (3) &  SIZE (4) &\\
(in p.e.)& (in $\%$)&( in $\%$) &( in $\%$) & ( in $\%$) &(in GeV)\\
\mr
$<$100&36&28& 31& 33 &30\\
(100,160)&10&11&12& 11& 50\\
(160,250)& 4& 6& 7& 5 &70\\
(250,400)& 3& 4& 4 &4 &100\\
$>$400& 1& 2& 3& 2 &150\\
\br
\end{tabular}
\end{indented}
\end{table} 

\subsection{Parameters for event selection}
One of the Hillas parameter that describes the image shape is the WIDTH, which is important for this analysis. Figure 5a shows a comparison of the WIDTH distributions of the simulated $\gamma$ (dotted histogram), false $\gamma$ (solid histogram) and one $\pi^0$ images (dashed histogram). All distributions are normalized to 1. The distribution of one $\pi^0$ events is  shifted a little bit towards larger WIDTHS in comparison to real $\gamma$-ray events, because the angular distribution of the charged particles in one $\pi^0$ events is wider than that of a $\gamma$ cascade. The separation angle between the decay products depends on the energy of the $\pi^0$ and how the energy is shared between the decay products.
This separation angle can be comparable to the opening angle of the Cherenkov light cone (for example a $\pi^0$ with an energy higher than 20 GeV may decay into two $\gamma$'s with such a separation angle).
The distribution of false $\gamma$ images is shifted towards lower WIDTHS in comparison to real $\gamma$-ray events. This may be explained by the fact that the detected false $\gamma$ events are subcascdes which begin deeper in the atmosphere and thus the observed cascade parts are younger and the angular distribution of electrons and positrons is narrower than in the primary $\gamma$ cascades (see e.g. \cite{giller}).\\
Another natural reason for the differences between the shown WIDTH distributions of $\gamma$-ray, false $\gamma$ and one $\pi^0$ events are the different OFF distributions. A comparison of the WIDTH distributions of the  $\gamma$-rays and both false $\gamma$ and one $\pi^0$ images with a small OFF angle (below 2$^o$) is presented in figure 5b. The differences seen in figure 5a have almost disappeared. The WIDTH is not a good variable to discriminate true $\gamma$-rays from 
false $\gamma$ and one $\pi^0$ events.\\
\begin{figure}
\begin{center}
\includegraphics*[width=14cm]{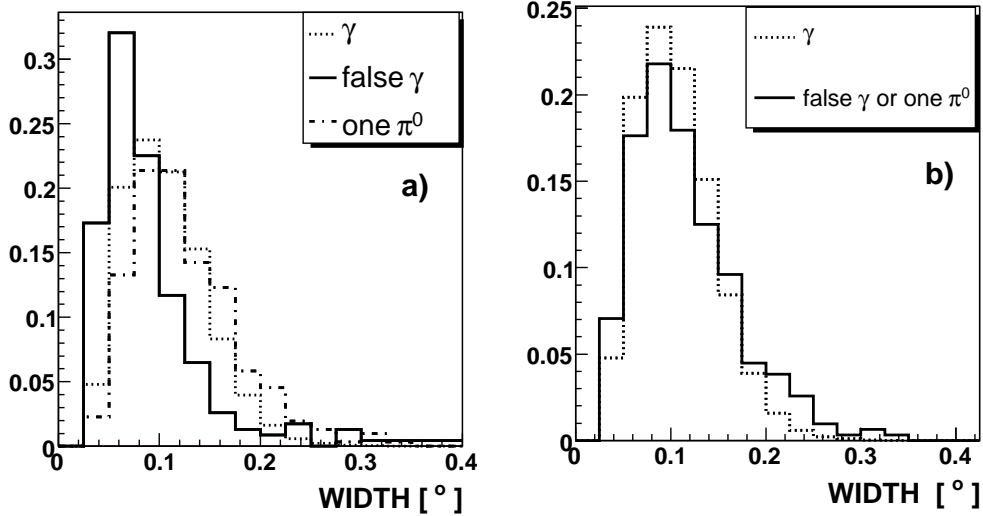}
\end{center}
\caption{{\bf a)} WIDTH distributions for true $\gamma$, false $\gamma$ and one $\pi^0$ events, for a trigger threshold of 3 p.e.; {\bf b)} The same as in a) but with an OFF angle below 2$^o$. All histograms are normalized to 1.}
\label{width}
\end{figure} 
The second Hillas parameter that is describing the image shape is LENGTH. Figure 6a shows the LENGTH distributions corresponding to true $\gamma$ (dotted), false $\gamma$ (solid) and one $\pi^0$ (dashed histogram) images. The distribution of true $\gamma$-ray is narrower than that of false $\gamma$ and one $\pi^0$ events because of different inclination angles of the shower. The primary $\gamma$-rays were simulated parallel to the telescope's axes and for geometrical reasons only a part of the longitudinal cascade is detected by the telescope. False $\gamma$ and one $\pi^0$ events are inclined to the telescope's axes and thus a longer part of the shower may be visible to the detector. This results in a larger image LENGTH. It is also possible that the observed longitudinal part of a  proton shower imitating a real $\gamma$-ray is shorter than a true $\gamma$-ray. One may expect that the false $\gamma$ image has a smaller length. In case of one $\pi^0$ events lower LENGTHS are not observed because the products of its decay are separated by a non negligible angle.\\
The LENGTH parameter is used in the $\gamma$/hadron separation, because it is correlated with the observed longitudinal part of the shower regardless of the primary particle type and thus connected with the shower direction.\\
A comparison of LENGTH distributions for primary $\gamma$-rays and both false $\gamma$ and one $\pi^0$ events with a small OFF angle (below 2$^o$) is presented in figure 6b. The differences seen in figure 6a have almost disappeared at low LENGTHS because the arrival direction distributions of proton showers and primary $\gamma$ cascades are now more similar. There are still images with larger LENGTH compared to electromagnetic cascades. This is caused by the separation angle of the $\pi^0$ decay products.\\
\begin{figure}
\begin{center}
\includegraphics*[width=14cm]{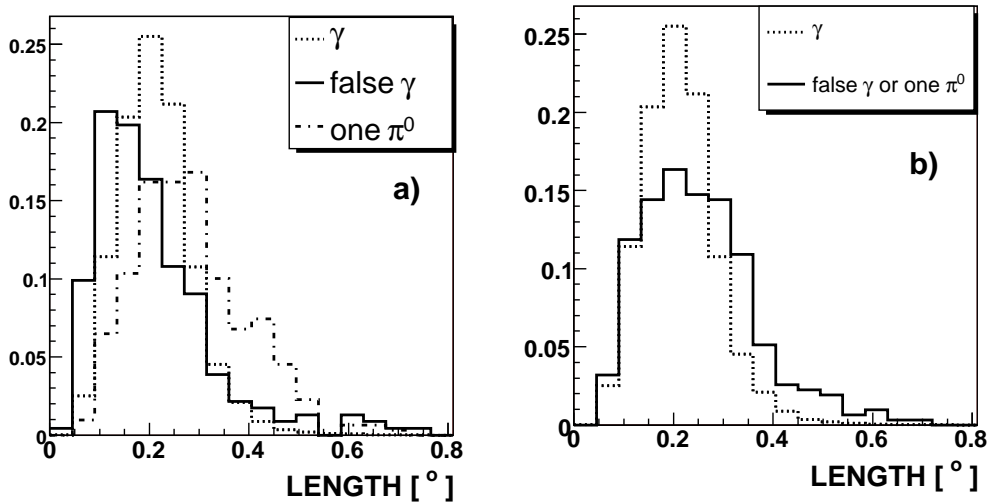}
\end{center}
\caption{{\bf a)} LENGTH distributions for true $\gamma$, false $\gamma$ and one $\pi^0$ events, for a trigger threshold of 3 p.e.; {\bf b)} The same as in a) but with an OFF angle below 2$^o$. All histograms are normalized to 1.}
\label{length}
\end{figure} 

\subsection{Ratio of the hardly reducible background to the signal from Crab Nebulae} 
The $\theta^2$ parameter is used to determine the direction of the primary particle in a stereo system of IACTs. $\theta^2$ is the square of the angular distance between the reconstructed direction of the shower and the $\gamma$-ray source position in the camera. The fixed direction of the simulated $\gamma$-rays results in a very steep $\theta^2$ distribution that peaks at 0. Proton initiated showers feature a flat $\theta^2$ distribution. 
I have checked that the $\theta^2$ distributions of electromagnetic, one $\pi^0$ and false $\gamma$ events have the same shape as the whole proton sample.
 $\gamma$/hadron separation based on the $\theta^2$ parameter is not affected by the detection of protonic events imitating  true $\gamma$-rays. However the efficiency of the final $\gamma$ selection (using $\theta^2$) depends on the chosen SIZE interval because the fraction of one $\pi^0$ or false $\gamma$ images is SIZE dependent.\\
The expected number of both false $\gamma$ and one $\pi^0$ images with $\theta^2 < 0.02^{o^{2}}$ was estimated using the primary proton and He spectra \cite{bess,hareyama}. The background from (not simulated) primary He has been added assuming that they produce the same fraction of hardly reducible background as primary protons. Accordingly the number of irreducible background events increases by 5.5$\%$. The Crab spectrum measured by MAGIC in 2005 \cite{wagner} was used to calculate the number of real $\gamma$ events which are expected within the same time interval as the cosmic ray background. The spectrum was extrapolated to energies below 300 GeV. New measurements of the spectrum show that it is flatter below 300 GeV \cite{crab2008} thus the expected number of true $\gamma$-rays can be overestimated. No additional $\gamma$/hadron separation (except $\theta^2$) has been applied to the simulations before the calculation of the expected numbers of true $\gamma$-rays and hadronic background.  
The ratio of both false $\gamma$ and one $\pi^0$ images to that of primary $\gamma$-rays from the Crab Nebula direction has been calculated in different SIZE bins. Figure 7a shows the result for trigger thresholds 2, 3, 4 and 5 p.e., respectively.  The event was classified into a SIZE bin if SIZE detected by a fixed telescope was in the chosen range - condition (3) of SIZE, which is described in the text above. Similar values were obtained for conditions (1) and (2).\\
The presented ratio decreases from 1.8 to 0.6 when the trigger threshold increases from 2 p.e. to 5 p.e. for SIZES below 100 p.e. 
The ratio decreases with larger SIZES for all trigger thresholds.
For low SIZES the presented ratio depends strongly on the trigger threshold, while at larger SIZES the dependence is less pronounced. It can be seen in figure 7a that the $\gamma$/hadron separation is much more difficult for images with SIZES $<$ 100 p.e. due to the occurrence of false $\gamma$ and one $\pi^0$ images.\\
A similar ratio has been shown in \cite{sob2007} for a single telescope with the same mirror area. A comparison between  a single IACT and a stereo system can be done for trigger thresholds of 4 p.e. (single) and 3 p.e. (stereo) because in both cases the same energy threshold has been obtained. The ratio of the expected number of both false $\gamma$ and one $\pi^0$ images to that of primary $\gamma$-rays is approximately eighth times lower for the stereo system than for a single telescope. Using the stereo imaging technique measurement one may expect a significant reduction of this kind of background in comparison to a single telescope.\\
An additional background reduction is expected if a $\gamma$/hadron separation method is applied before cutting on the $\theta^2$ parameter. Figure 7b presents the same ratio as figure 7a but calculated after a simple $\gamma$-ray selection -  cuts in WIDTH and LENGTH ($0.05^o<$ WIDTH $<0.15^o$ and $0.1^o <$ LENGTH $<0.3^o$) have been applied. No false $\gamma$ and one $\pi^0$ events with SIZE larger than 250 p.e. were found after those cuts in the simulated Monte Carlo data-set. The detection of one true $\gamma$-ray with SIZE below 100 p.e. is associated with the detection of about one false $\gamma$ or one $\pi^0$ event, which survive the WIDTH and LENGTH cuts. 
There are more sophisticated methods for the $\gamma$/ hadron separation \cite{random2008} which are more efficient in the background reduction.
 However the similarity of true and false $\gamma$ image shapes results also in a significant reduction of the number of the separated true $\gamma$-rays in the low SIZE range. This is one of the main reasons why the sensitivity of IACTs is worsening in the low energy range. The sensitivity at lower energies is also reduced due to the high fluctuations of the shower development \cite{sob2009}.\\
\begin{figure}
\begin{center}
\includegraphics*[width=14cm]{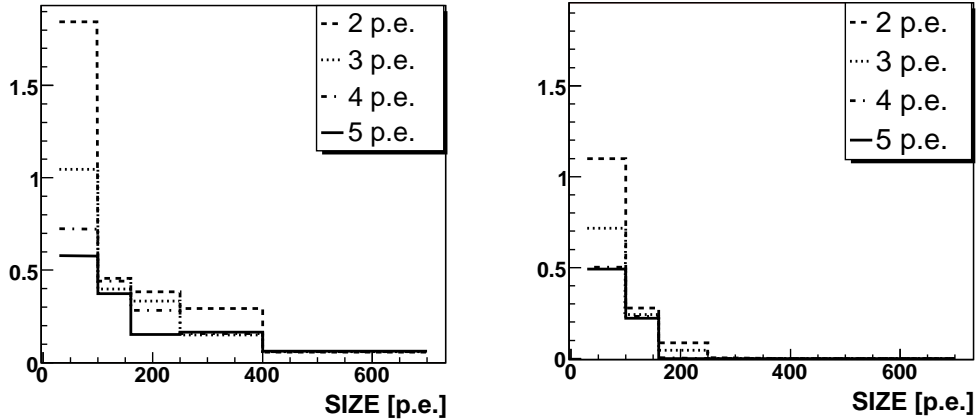}
\end{center}
\caption{Ratio of the expected number of false $\gamma$ and one $\pi^0$ images to the expected number of true $\gamma$-rays (from the direction of the Crab Nebula) in different SIZE bins for $\theta^2$ smaller than 0.02: {\bf a)}before WIDTH and LENGTH cuts; {\bf b)}after simple WIDTH and LENGTH cuts (see text). Different line styles correspond to different trigger thresholds.}
\label{fralpha}
\end{figure} 

\section {Conclusions}
The simulations presented in this paper showed that two large Cherenkov telescopes working in stereo mode may be triggered by photons from one electromagnetic subcascade of proton-induced showers. The occurrence of this kind of background is mostly caused by low energy proton induced showers - approximately only $5\%$ of such events have a primary energy above 200 GeV for a trigger threshold 3 of p.e. The distance between the center of the telescope system and the shower axis  may be very large. No specific range of the inclination angle of the shower to the telescopes axes has been found, as expected.  
The proton background contains 13$\%$ of false $\gamma$ events at a trigger threshold of 2 p.e. which decreases to 3$\%$ at a trigger threshold of 5 p.e. 
The false $\gamma$ images are concentrated at small SIZE values.
 The images of the false and true $\gamma$-ray have similar shapes - the differences in the WIDTH and LENGTH distributions are mostly caused by different shower direction distributions.\\
The images from two electromagnetic subcascades from the same $\pi^0$ decay may also be registered by a stereo system of large IACTs. 92$\%$ of such events correspond to protons with a primary energy of protons lower than 200 GeV for a trigger threshold of 3 p.e. The fraction of such events in the total proton background decreases with increasing trigger threshold from 12$\%$ to 6$\%$ for thresholds from 2 to 5 p.e. One $\pi^0$ images are a little wider and longer than true $\gamma$ images. The differences are smaller if the direction of the background events is limited.\\
It has been shown in \cite{sob2007} that efficient $\gamma$-ray separation from one $\pi^0$ and false $\gamma$ images is not possible by using only the shape parameters of a single MAGIC-like telescope. The results presented in this paper show that a system of two large IACTs such as the MAGIC II experiment faces similar problems. 
It was suggested in \cite{maier} that hadronic showers in which most of the primary energy is transferred to electromagnetic subcascades during the first few interactions constitute a hardly reducible background. The quantities estimated in this paper depend on the interaction models, which were chosen in the extensive air shower simulation. It has been shown in \cite{maier} that the GHEISHA model gives the lowest probability that the majority of the primary energy is deposited in the electromagnetic part of the shower. This model was used in simulations presented in this paper. Thus other models may produce an even higher fraction of false $\gamma$ and one $\pi^0$ events than calculated and shown in  table 2. The number of irreducible background events may be estimated more accurately by the simulation of heavier nuclei. However, since the flux of heavier nuclei is much lower than that of the proton initiated showers \cite{hareyama}, the effect on the total number of irreducible background events should be relatively small.\\
The  $\gamma$/hadron separation is much more difficult due to the existence of the false $\gamma$ and one $\pi^0$ images in the low energy region.\\ 
The ratio of the number of the expected background from false $\gamma$ and one $\pi^0$ events to the number of the triggered high energy photons from the direction of the Crab Nebula is a few times lower for a stereo measurement than for a single telescope detection \cite{sob2007} in the same energy range.
This natural background suppression is caused by a more precise estimation of the shower direction for two IACTs ($\theta^2 <0.02$) compared to a single Cherenkov telescope (ALPHA $<15^o$). However a significant number of the false $\gamma$ and one $\pi^0$ events survive a simple selection based on the image shape parameters for SIZES lower than 250 p.e. (what corresponds to a primary energy below 100 GeV). In conclusion the sensitivity of a stereo system of IACTs is worsening at lower energies because of the existence of a hardly reducible background from false $\gamma$ and one $\pi^0$ events.

\ack
This work was supported by the Polish KBN grant No. N N203390834.

\end{document}